\documentclass[twocolumn]{svjour3}          

\smartqed  
\usepackage{amsmath}
\usepackage{eucal}
\usepackage{amsfonts}
\usepackage{graphicx}
\journalname{arXiv}
\begin{document}

\title{Three fundamental problems in risk modelling on big data: an information theory view 
}


\author{Jiamin Yu        
}


\institute{Jiamin Yu 
		\at Shanghai Lixin University of Accounting and Finance, \\
			Shangchun Road 995, Pudong, Shanghai 201209, China  \\
              \email{iambabyface@hotmail.com}           
}

\date{Received: date / Accepted: date}

\maketitle

\begin{abstract}
Since Claude Shannon founded Information Theory, information theory has widely fostered other scientific fields, such as statistics, artificial intelligence, biology, behavioural science, neuroscience, economics, and finance. Unfortunately, actuarial science has hardly benefited from information theory. So far, only one actuarial paper on information theory can be searched by academic search engines. Undoubtedly, information and risk, both as Uncertainty, are constrained by entropy’s law. Today's insurance big data era means more data and more information. It is unacceptable for risk management and actuarial science to ignore information theory. Therefore, this paper aims to exploit information theory to discover performance limits of insurance big data systems and seek guidance for risk modelling and the development of actuarial pricing systems.
\end{abstract}

\section{Introduction}
\label{intro}
Since Claude Shannon published his most important paper “{\itshape A mathematical theory of communication}” in 1948, {\itshape Information Theory} has been established and widely fostered many scientific fields, such as statistics, artificial intelligence, biology, behavioral science, neuroscience, economics, and finance. Even in finance, which is most similar to actuarial science, for example, there is a lot of studies applying information theory, such as the financial value of information \cite{barron1988a}, portfolio theory \cite{ambachtsheer2005beyond}, credit ratings \cite{bariviera2015efficiency}, and the impact of social media on financial markets \cite{zheludev2015when}. In contrast, there is only one paper on information theory in actuarial science so far. Reference \cite{sachlas2014residual} discussed the residual entropy and past entropy in actuarial science and survival models for the deductible and limit of insurance claims. Unfortunately, from the search results of academic search engines, it can be said that actuarial science has hardly benefited from information theory.
\paragraph{}
First of all, theoretical research of risk theory and actuarial science requires information theory. By definition, both {\itshape risk} (in risk theory) and  {\itshape information} (in information theory) are uncertainties theoretically. In the universe, everything is energy, and information is also energy. All spontaneous energy evolution processes, including risk processes, obey the second law of thermodynamics, also known as  {\itshape the Principle of Entropy Increase}. Undoubtedly, information and risk, both as uncertainties, are constrained by information entropy’s law. Obviously, information theory also applies to risk theory and actuarial science. 
\paragraph{}
Secondly, the practical application of risk management in the insurance industry (such as self-driving car insurance, cyber insurance etc.) also requires information theory. Today's insurance industry has entered an era of insurance big data. This means that more huge amounts of data (i.e. information) can be used for actuarial modeling and risk management. Faced with massive information of big data, it is unacceptable for risk management and actuarial science to continue to ignore information theory.
\paragraph{}
Therefore, in order to meet the above theoretical and practical requirements, this paper aims to exploit information theory to discover the performance limits of insurance big data systems, and seek guidance for risk modeling and the development of actuarial pricing systems.
\paragraph{}
The rest of this paper is organized as follows. Setion~\ref{sec:1} presents a novel communication system framework of risk model, which promotes the actuarial model to achieve both more risk classification and less pricing error. Section 3 ascribes risk classifications to a refinement problem of risk decoding, which can be solved by refinement of risk output event. Section 4 shows that the mutual information between risk input and output determines the performance of the actuarial model, and deterministic or causal model is the optimal actuarial model. Section 5 proposes that risk big data might cause risk information distortion, and lossless data collection design can significantly improve the performance of the actuarial model. Conclusions section summarizes main conclusions of this research and discusses future research outlook.

\section{Communication system framework of risk model}
\label{sec:1}
\subsection{Traditional risk model}
\label{sec:11}
The traditional actuarial risk model is the currently popular experience rating model, which predicts future losses (net risk premium) based on the historical loss (claim) experience of the insured subject. The experience rating model can be abstracted into the schematic diagram as below.
\begin{figure}[h]
\centering
  \includegraphics[width=0.45\textwidth]{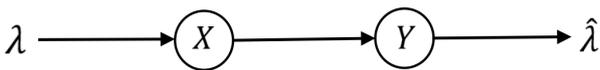}
\caption{The traditional actuarial risk model schematic.}
\label{fig:1}       
\end{figure}
\\Where, $\lambda$ is the real world risk level(unknown hidden variable) and $\hat{\lambda}$ is the posterior estimated risk; $X$ and $Y$ are the underwriting risk variables (rate factors) and risk event (e.g. accident loss record) of the insured subject. The core mathematical problem is to solve the conditional probability distribution of the risk level $\lambda$ under the observed conditions of $X$ and $Y$:
\begin{equation}
P(\lambda|X,Y) \notag
\end{equation}
\paragraph{}
The experience rating risk model implies a time-invariant assumption, that is, the future $\lambda$ and historical $\lambda$ remain unchanged. This obviously contradicts to our common sense, for instance, the risk of traffic accidents during rush hours is much higher than slack hours. It can be seen that traditional risk models cannot meet the requirements of precise pricing of big data.
\subsection{Communication system framework of risk model}
\label{sec:12}
\subsubsection{Time-varying risk assumption}
In the real world, risk changes over time generally. Assuming that the risk in Fig.~\ref{fig:1} is time variant, the risk model becomes as follows
\begin{figure}[h]
\centering
  \includegraphics[width=0.45\textwidth]{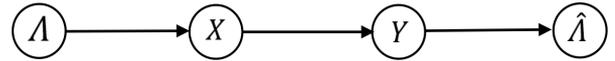}
\caption{The time-varying risk model schematic.}
\label{fig:2}       
\end{figure}
\\Where, the original $\lambda$ in Fig.~\ref{fig:1} is replaced to $\Lambda$, which represents the time-varying risk level, and $\hat{\Lambda}$ is the corresponding estimated risk level. Mathematically, $\Lambda$ can be regarded as a function $\Lambda(t)$ of time $t$, or even a stochastic process $\Lambda_t$.
\subsubsection{Risk informationization}
\begin{figure*}
\centering
  \includegraphics[width=0.95\textwidth]{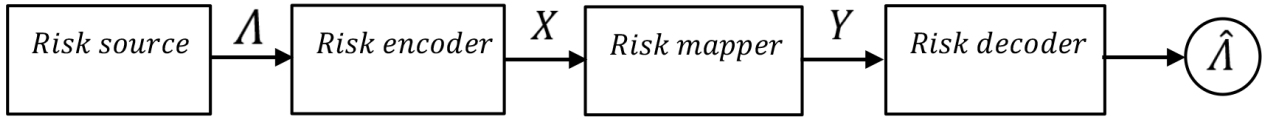}
\caption{The communication system framework of time-varying risk model.}
\label{fig:3}       
\end{figure*}

According to the Merriam-Webster Dictionary, {\itshape risk} is {\itshape possibility of loss or injury}. Following the definition of {\itshape information measurement} in Shannon Information Theory \cite{moser2012a}, the {\itshape information measurement of risk} is the information (or uncertainty) measurement of loss or injury. Consequently, the actuarial problem in Fig.~\ref{fig:2} becomes the information problem of the communication system in Fig.~\ref{fig:3} as below.
\paragraph{Paragraph headings} Use paragraph headings as needed.
\paragraph{}
As illustrated in Fig.~\ref{fig:3}, the original actuarial estimation process(from $\Lambda$ to  $\hat{\Lambda}$) in Fig.~\ref{fig:2} is transformed into the risk information transmission process(from $\Lambda$ to  $\hat{\Lambda}$) in Fig.~\ref{fig:3}. Where, $\Lambda$ originates from real-world {\itshape risk source}; With the {\itshape risk encoder}, the unknown risk $\Lambda$ is collected as the risk input $X$ by big data acquisition; $X$ generates risk event $Y$ through the {\itshape risk mapper}; Relying on the {\itshape risk decoder}, the event $Y$ can be decoded into risk estimate $\hat{\Lambda}$ finally. Further, this risk system is decomposed into three simple input-output subsystems, namely risk encoder, risk mapper, and risk decoder.
\subsection{Equivalence of actuarial problems: information problems}
\label{sec:13}
In the context of insurance big data, the actuarial model adds two new requirements. One is more risk classification, the other is higher pricing accuracy. If the set of various risk levels is represented as set $\mathcal{C}$, the first requirement of risk classification can be described as the following mathematical problem
\begin{equation}
\max\big(|\mathcal{C}|\big) \notag
\end{equation}
Where, $|\mathcal{C}|$ is the number of risk levels in the risk classification set $\mathcal{C}$.  Element $c$ represents a certain risk level in set $\mathcal{C}$. The probability of pricing errors with risk level $c$, can be defined as follows
\begin{equation}
P^e_c=P(\hat{\Lambda}_c \neq \Lambda_c),\quad c \in \mathcal{C} \label{eq:1}
\end{equation}
Where, $\Lambda_c$ and $\hat{\Lambda}_c$ are the real and estimated risk of $c$ respectively. Similarly, the second requirement for precise pricing can be described as the following mathematical problem
\begin{equation}
\min_{c \in \mathcal{C}}(P^e_c) \notag
\end{equation}
\paragraph{}
However, it seems that two mathematical problems $\max(|\mathcal{C}|)$ and $\min_{c \in \mathcal{C}}(P^e_c)$ are paradoxical: In case of more risk classification, it means $\hat{\Lambda}_c$ contains more uncertainty; In case of more risk classification precise pricing, it implies $\hat{\Lambda}_c$ has less uncertainty. This paradox is consistent with the {\it aggregation} operation in experience rating: {\it aggregation can reduce pricing errors but decay risk classification meanwhile}.
\paragraph{}
Once the actuarial risk model is modeled as a risk communication system, the traditional actuarial problem can be transformed into one problem of information transmission from real-world risk $\Lambda$ to estimated risk $\hat{\Lambda}$. Thus, according to the {\it mutual information theorem} of information theory \cite{cover2006elements}, the optimal risk modeling criterion is directly inferred to maximize {\it the mutual information between real and estimated risk}. According to the idea of information theory, the above two mathematical problems can be transformed into the following information expressions
\begin{equation}
\max_{\mathcal{C}}\big(H(\hat{\Lambda})\big)\quad and \quad \max_{\mathcal{C}}\big(I(\Lambda,\hat{\Lambda})\big)\notag
\end{equation}
\\Where, $H(\hat{\Lambda})$ is the information entropy of $\hat{\Lambda}$, and $I(\Lambda,\hat{\Lambda})$ is the mutual information between $\Lambda$ and $\hat{\Lambda}$. $\max_{\mathcal{C}}(H(\hat{\Lambda}))$ is equivalent to $\max(|\mathcal{C}|)$ (more uncertainty means greater entropy information entropy), and $max_{\mathcal{C}}(I(\Lambda,\hat{\Lambda}))$ is equivalent to $\min_{c \in \mathcal{C}}(P^e_c)$ (less uncertainty means greater mutual information). Thus, information theory eliminates the paradox caused by the two requirements and creates the achievability of the optimal actuarial model.
\paragraph{}
Further, according to Fig.~\ref{fig:3}, the whole risk information transmission system is composed of risk encoder, risk mapper, and risk decoder three subsystems. Consequently, the information problems of the whole risk communication system can be decomposed into the information problems of three subsystems. The following sections will discuss these information issues in detail.

\section{Refinement problem of risk decoding}
\label{sec:2}
\subsection{Information entropy of risk decoding process}
Next, focus on the {\it risk decoder} in Fig.~\ref{fig:3}, an input and output systems shown as below.
\begin{figure}[h]
\centering
  \includegraphics[width=0.3\textwidth]{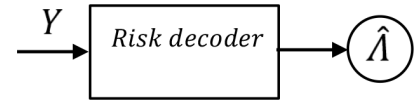}
\caption{The input-output system of risk decoder.}
\label{fig:4}       
\end{figure}
According to the lemma of mutual information transmission (i.e. Data processing inequality) \cite{moser2012a}, the following inequality is obtained
\begin{equation}
I(\Lambda,\hat{\Lambda}) \le I(Y,\hat{\Lambda}) \le \min\bigl\{H(\hat{\Lambda}),H(Y)\bigr\}
\label{eq:2}
\end{equation}
Mathematically, inequality (\ref{eq:2}) reveals that, maximization of $\max_{\mathcal{C}}(I(\Lambda,\hat{\Lambda}))$ depends on the maximization of $I(Y,\hat{\Lambda})$, which prerequisite is to maximize $H(Y)$(i.e. the information entropy of risk event $Y$). Therefore, increasing the information entropy of risk event $Y$ is a necessary condition for achieving more risk classification.
\subsection{Solution: refinement of risk event}
\begin{figure*}[htbp]
\centering
  \includegraphics[width=0.8\textwidth]{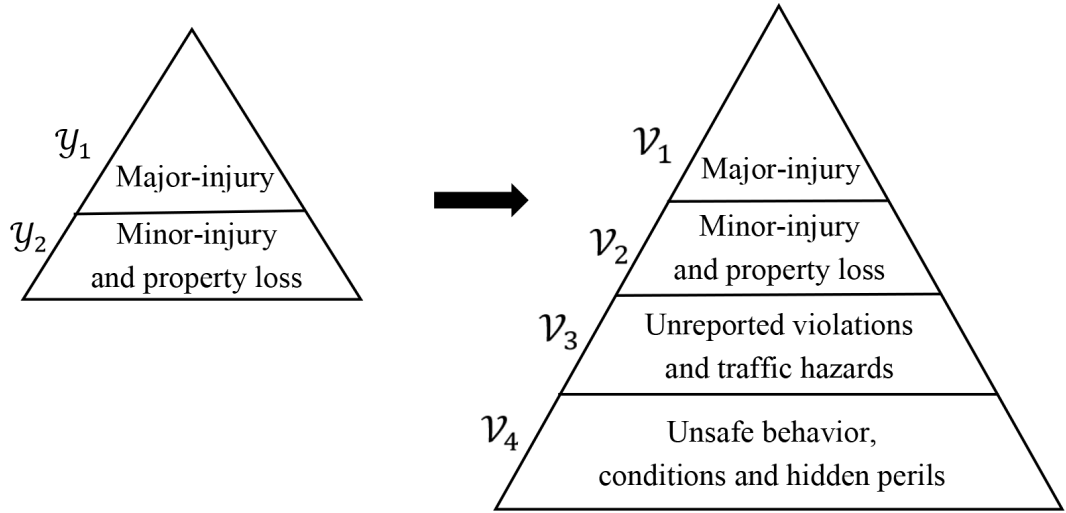}
\caption{The Heinrich’s Triangle of automobile insurance and 4- refinement of risk event.}
\label{fig:5}       
\end{figure*}

Let $\mathcal{Y}$ be the set of risk event $Y$, and divide $\mathcal{Y}$ into $k$ mutually disjoint sets, $\mathcal{V}_1,\mathcal{V}_2,\ldots,\mathcal{V}_k$, satisfying
\begin{equation}
\bigcup^k_{i=1}\mathcal{V}_i=\mathcal{Y}
\label{eq:3}
\end{equation}
Then $\mathcal{V}$ is called a $k$-refinement of risk event set $Y$. Apparently, it holds
\begin{equation}
H(Y) \le H(V) \quad and \quad I(Y,\hat{\Lambda}) \le I(V,\hat{\Lambda}) 
\label{eq:4}
\end{equation}
Equation (\ref{eq:4}) shows that, {\it it is mathematically feasible to refine risk event $V$ for more risk classifications}.
\paragraph{}
Fortunately, the Heinrich’s Safety Triangle(Heinrich’s Law) \cite{heinrich1959industrial} also ensures that the refinement of risk event has strong operability. Taking car insurance as an example, autonomous driving big data can refine risk event $Y$ as Fig.~\ref{fig:5}.
\paragraph{}
In Fig.~\ref{fig:5}, traditional car insurance can achieve a 2-refinement of risk event, $\mathcal{Y}_1$ ({\it Major-injury})  and $\mathcal{Y}_2$ ({\it Minor-injury/property loss}); With the aid of self-driving technology, insurance companies can achieve a 4-refinement of risk event, by appending $\mathcal{V}_3$ ({\it Unreported violations and traffic hazards}) and $\mathcal{V}_4$ ({\it Unsafe behavior, conditions and hidden perils}).
\paragraph{Significance}
The essence of risk event refinement is to ensure that {\it risk events ($V$) contain more risk information redundancy than estimated risks ($\hat{\Lambda}$), so as to improve the capacity of risk decoding}.

\section{Capacity problem of risk mapping}
\label{sec:3}
\subsection{Mutual information of risk mapping process}
\begin{figure}[h]
\centering
  \includegraphics[width=0.3\textwidth]{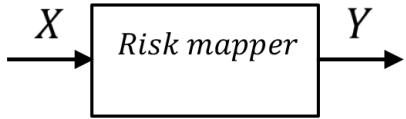}
\caption{The input-output system of risk mapper.}
\label{fig:6}       
\end{figure}

Next look at the {\it risk mapper} in Figure \ref{fig:3}, an input and output system shown as Fig.~\ref{fig:6}.
\paragraph{}
Just like inequality (\ref{eq:2}), there is also a data processing inequality as follow
\begin{equation}
I(\Lambda,\hat{\Lambda}) \le I(X,Y) \le \min\bigl\{H(X),H(Y)\bigr\}
\label{eq:5}
\end{equation}
Where, $X$ and $Y$ are the risk input and output of risk mapper, and $I(X,Y)$ is the mutual information between $X$ and $Y$. $I(X,Y)$ describes the reliability (or shared information) that risk input variables ($X$) generate (mapping) risk events ($Y$). For a noisy (or erroneous) risk model, {\it Risk Mapping Capacity} is the maximum reliable mapping rate, which depends on the mutual information $I(X,Y)$ and information entropy $H(X)$ (inequality (\ref{eq:5}) gives these bounds). Therefore, the design of a good risk model should relate to optimize $I(X,Y)$ and $H(X)$.

\subsection{Solution: maximization of risk mapping capacity}
According to Shannon's Channel Coding Theorem \cite{cover2006elements}\cite{moser2012a}, the limit of the reliable risk mapping rate is the maximum mutual information $I(X,Y)$ (i.e. the "common uncertainty" of $X$ and $Y$). Thus, entropy and mutual information, can be used to determine the optimality of an actuarial system.
\paragraph{}
According to monotonicity and convexity of mutual information, $I(X,Y)$ is a convex function of $P(Y|X)$ for fixed $P(X)$ (which means that the actuary cannot change the pricing variables $X$ except for the actuarial model $P(Y|X)$). It can be derived from its convexity as
\begin{equation}
\lim_{P(Y|X) \rightarrow 1}I(X,Y)=\sup\bigl(I(X,Y)\bigr)
\label{eq:6}
\end{equation}
Where $\sup(I(X,Y))$ is the supremum of $I(X,Y)$. Equation (\ref{eq:6}) implies that when $X$ and $Y$ are close to one-to-one mapping, mutual information $I(X,Y)$  tends to achieve the maximum value.
\paragraph{}
According to another property of mutual information, $I(X,Y)\le H(Y)$ with equality holding iff $Y$ is a function of $X$. That is
\begin{equation}
Y=G(X) \Rightarrow I(X,Y)=H(Y)
\label{eq:7}
\end{equation}
Combining equation (\ref{eq:6}) and (\ref{eq:7}), it shows that deterministic risk generating model $G(X)$ will cause
\begin{equation}
I(X,Y)=H(Y)=\sup\bigl(I(X,Y)\bigr)
\label{eq:8}
\end{equation}
In causal inference theory, $G(X)$ is called {\it structural equation} by \cite{pearl2018theoretical}.
\paragraph{}
Taking rear-end collision risk as an example, according to the vehicle dynamic model in reference \cite{brill1972a}, let
\begin{equation}
y=v_2 h_2+\frac{v^2_1}{2a_1}-v_2 r_2-\frac{v^2_2}{2a_2}
\label{eq:9}
\end{equation}
Where, $v_1$ and $a_1$ represent the initial speed and braking deceleration of the preceding vehicle 1, $v_2$ and $a_2$ are the initial speed and braking deceleration of the following vehicle 2, and $h_2$ and $r_2$ represent the time headway (gap) and reaction time of the following vehicle driver; $y$ is the risk event outcome variable. Here $v_1$, $v_2$, $a_1$, $a_2$, $h_2$,and $r_2$ belong to the risk input ($X$), and $y$ belongs to the risk output ($Y$). 
\paragraph{}
Under the condition that $X$ is known, whether  a collision  occurs can be judged directly based on the calculation result of $Y$ ($y>0$: no collision; $y \le 0$: collision). Apparently the actuarial dynamics models similar to (\ref{eq:9}) can satisfy maximization of risk mapping capacity, because of their one-to-one mapping.
\paragraph{Significance}
Maximization of actuarial model’s risk mapping capacity is to create deterministic or causal risk generating model.

\section{Distortion problem of risk variables}
\label{sec:4}
\subsection{Information entropy of big data risk variables}
Finally look at the {\it risk encoder} in Fig.~\ref{fig:3}, an input and output system shown as below
\begin{figure}[h]
\centering
  \includegraphics[width=0.45\textwidth]{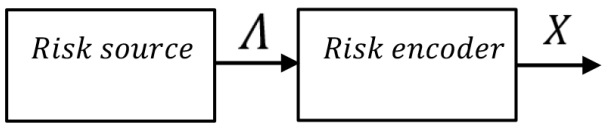}
\caption{The input-output system of risk encoder.}
\label{fig:7}       
\end{figure}
As shown in Fig.~\ref{fig:7}, the role of the risk encoder is to sample real-world risks ($\Lambda$) as big data risk input variables ($X$). Risk variables are usually continuous time variables, and $X$ are discrete time variables. According to the Nyquist-Shannon sampling theorem \cite{cover2006elements}: Discrete sampling sequence may have information loss. Therefore, how to find the less-distortion and compact representation of real world risk information? This is the essential problem of risk encoder, which is directly related to the design of data collection on big data.

\subsection{Constraints of risk big data design: multi-objective optimization}
According to Shannon's Source Coding Theorem \cite{cover2006elements}\cite{moser2012a}, insurance big data variables ($X$) should be designed with redundant information to carry the information entropy of real risk as lossless as possible. This can be expressed mathematically as
\begin{equation}
H(X)> H(\Lambda) \quad or \quad H(X)> H(\hat{\Lambda})
\label{eq:10}
\end{equation}
Where, $H(X)>H(\Lambda)$ is a strong constraint, and $H(X)>H(\hat{\Lambda})$ is a weak constraint, because $\Lambda$ is usually unknown.
\paragraph{}
In practice, the cost of big data collection is unavoidable. Therefore, (\ref{eq:10}) must consider the cost constraints, namely
\begin{equation}
\max_{H(X)> H(\hat{\Lambda},Cost(X)\le CoD}H(X)
\label{eq:11}
\end{equation}
Where, $Cost(X)$ is the cost function of big data collection ($X$), and $CoD$ is the upper bound of cost.
\paragraph{}
In addition, it is worth emphasizing that the big data design of ($X$) will directly affect the risk mapping capacity of the actuarial model (i.e. {\it risk mapper}). According to monotonicity and convexity of mutual information, $I(X,Y)$ is a concave function of  $P(X)$ for fixed $P(Y|X)$ (which means the actuarial model $P(Y|X)$ unchanged). It can be derived from its concavity as
\begin{equation}
\max_{P(X)}H(X)=\sup{I(X,Y)}
\label{eq:12}
\end{equation}
Equation (\ref{eq:12}) implies the fact that, not all insurance big data can be modeled as an excellent actuarial pricing model, only {\it the big data collection design optimized by mutual information can realize the best performance of the actuarial model}.
\paragraph{}
(\ref{eq:11}) and (\ref{eq:12}) show that, the insurance big data design of ($X$) must not only extract the real-world risk characteristics without distortion, but also help the actuary find a good enough actuarial model.
\paragraph{}
Reference \cite{shinar2019crash} presents an interesting and extreme example, assuming that driving risk (safety) model does not consider human factors such as driver’s distraction, and only depends on the frequencies of wearing shoes observation variables in accident samples, it is easy to conclude: “Wearing shoes is the inevitable cause of the crash, because almost all drivers were wearing shoes when crashes occurred.” If the probability of driver wearing shoes is 99.999\%, substituting the probability $P(X)$  into (\ref{eq:12}), there is
\begin{equation}
\lim_{P(1)=0.99999,P(0)=0.00001}I(X,Y)\rightarrow 0 \ll\sup{I(X,Y)}
\label{eq:13}
\end{equation}
According to $I(X,Y)$’s concavity, it can be directly inferred that the risk variable of wearing shoes is a very poor risk characteristic variable, so it is impossible to model it as a good actuarial model.
\section{Conclusions}
The traditional actuarial model is prone to cause ecological paradox and Simpson's paradox due to their correlation statistical assumptions, which were reported in Reference \cite{davis2004possible}, a research on traffic big data. In this paper, the big data risk model is tackled as a risk information input-output system. This novel modeling framework can apply well-developed information theory to simultaneously achieve more risk classifications and less pricing errors.
\paragraph{}
The most important contribution of this work is to provide a new perspective of big data actuarial modeling, which enables actuaries to use more sophisticated mathematical tools such as information theory and system dynamics for risk modeling. In particular, specific approaches for optimizing the performance of big data actuarial models are proposed, including big data collection design, risk event refinement, and even causal (deterministic) modelling.
\paragraph{}
The work in this paper could have the potential to be widely used in insurance big data risk modeling, especially self-driving car insurance, cyber insurance.


%
%

\bibliographystyle{amsplain}  
\bibliography{bibrefs}   			

%
%

\end{document}